\magnification=\magstep1   
\nopagenumbers
\parindent=12pt
\parskip=3pt
\baselineskip=24pt

\def\qf{q_{_F}}
\def\lf{\ell_{_F}}
\def\lya{Ly$\alpha$~forest~}
\def\lyad{Ly$\alpha$~}
\def\cool{{\cal L}}

\def\etal{{\it et~al.\ }}

\def\ie{{\it i.e.,~}}
\def\de{\partial}
\def\ltsima{$\; \buildrel < \over \sim \;$}
\def\simlt{\lower.5ex\hbox{\ltsima}}
\def\gtsima{$\; \buildrel > \over \sim \;$}
\def\simgt{\lower.5ex\hbox{\gtsima}}
\def\ref{\noindent\hangindent.5in\hangafter=1}
 
\centerline{ } 
\vskip 8truepc
\centerline{\bf  COSMOLOGICAL CONDUCTIVE/COOLING FRONTS}
\centerline{\bf  AS \lyad FOREST CLOUDS} 
\vskip 5truepc
\centerline{\bf A. Ferrara$^{1}$ {\rm and} Yu. Shchekinov$^{2}$}
\centerline{\it ${^1}$ Osservatorio Astrofisico di Arcetri,
Largo E. Fermi, 5, 50125 Florence, Italy}
\centerline{\it ${^2}$ Institute of Physics, Rostov University,
194 Stachki, 344104 Rostov on Don, Russia}
\vfill\eject
\vskip 5truecm   
\centerline{\bf ABSTRACT}
\medskip
We propose a simple model for the origin and evolution of \lya clouds
based on cosmological conductive/cooling fronts. In this model the \lya 
arises in the interfaces between the IGM and cold clouds that 
could be tentatively identified with protogalaxies. 
Most of the properties of the \lya absorbers are reproduced with a very 
restricted number of assumptions. Among these are the correct range of
HI column density, cloud sizes and redshift and HI column density distributions
for the absorbers. Several predictions and implications of the model are 
briefly discussed.

\vskip 2truecm
{\tt Subject Headings:} intergalactic medium --- quasars: absorption lines 
\vfill\eject
 
\centerline{\bf 1. Introduction}
\medskip
The remarkable observational progresses made in the last few years allow to
tackle the problem of the origin and evolution of \lya clouds in a more
firm and quantitative manner. 
In brief, the available observational constraints that every model for the \lya
clouds should
fulfill are the following. The typical range of hydrogen column density
is $10^{12}{\rm cm}^{-2} \le N_{HI} \le 10^{16}$~cm$^{-2}$         
(Dobrzycki \& Bechtold 1996).
Doppler parameters are typically $b=30$~km~s$^{-1}$, even if some claims
of much smaller values of $b$ have been made (Pettini \etal 1990; Carswell
\etal 1991; Rauch \etal 1993; Giallongo \etal 1993). Lower limits on the sizes are 
provided by a small number of  experiments using quasar pairs 
(Dinshaw \etal 1994, Bechtold \etal 1994) which generally agree that transverse 
sizes should be $\ge 100$~kpc.

Here we propose an alternative view of the origin and evolution of \lya
absorbers, based on the theory of conductive/cooling fronts described by
Ferrara \& Shchekinov (1993) (FS93). The basic idea is that \lya clouds may represent
the interface region between the IGM and cold clouds that could be tentatively
identified with protogalaxies. 
Since the IGM is in a thermally unstable state, the joint action of thermal conduction
and radiative instabilities drives {\it a cooling wave} into the hot IGM.
Thus, contrary to the common view in which thermal conduction should eventually 
lead to the evaporation and destruction of the cold phase, we argue that 
protogalaxies initiate the condensation process of gas which can be subsequently
accreted by the galaxy itself. The relatively rarefied transition layer between 
the two phases represents a suitable environment for the formation of the \lya forest
absorption lines. Although very simple, the model is able to reproduce most of
the properties of \lya clouds with a very restricted number of assumptions.
\bigskip
\centerline{\bf 2. Conductive/Cooling Fronts } 
\medskip
In this Section we describe briefly some results concerning conductive/cooling
(CC) fronts following FS93 and Ferrara \&
Shchekinov (1996). We define a CC front as the interface between 
a hot and cold gas phase, whose structure is governed by the combined effects of
thermal conduction and radiative cooling. As shown by FS93, 
dynamical 
effects (\ie shocks) can be important at the initial stages of the front 
evolution when the conductive timescale is shorter than the dynamical timescale 
of the system. However, after this transient the evolution relaxes to a
quasi-isobaric regime. We will therefore consider only evolved stages for which
pressure equilibrium holds approximately. For the same reason, we consider
only "classical"  Spitzer conductivity since saturation effects are important only
in the initial evolutionary phases (FS93). We neglect magnetic fields, mostly
because their existence and strength at high redshift are far from 
being established.

In the isobaric approximation, the hydrodynamic equations describing a CC front
reduce to the energy equation, which could be cast in Lagrangian form as 

$${\de T\over \de t}+ c_p^{-1} {\cal L}- {\de \over \de q}\left[ \chi {\de T\over 
\de q}\right]=0,\eqno(2.1)$$

\noindent where $T$ is the gas temperature, $\cool$ is the net cooling rate 
per unit mass, $c_p$ is the specific heat at constant pressure; $\chi= 
\kappa \rho /c_p$, where $\kappa=\eta T^{5/2}$ is the classical thermal
conduction coefficient, is the reduced thermal conduction coefficient.  
The Lagrangian mass variable is 

$$ q = \int_{x_0}^x dx \rho(x,t)$$

\noindent where for all $t$ the coordinate of a ``reference'' particle $x_0(t)$
is determined by the condition $v(x_0,t)=0$. In eq. (2.1) lengths are normalized
to the Field length (Field 1965) (Lagrangian formulation), $\qf=(\chi T
c_p/\cool)^{1/2}$, and time to the cooling time $t_c=c_p T/\cool$, both calculated
for the hot medium.

The cooling function depends on the details of the microscopic processes
responsible for the heating and cooling of the gas; for our purposes, however,
it will be sufficient to consider a general form of $\cool$ reproducing
a two phase medium with a cold stable, and a hot unstable equilibrium.
This assumption closely resembles the IGM, supposedly constituted by 
cold clouds in pressure equilibrium with a surrounding, thermally unstable,
hot diffuse gas. The simplest analytical function that retains such 
characteristics is $\cool(T)=  T (1 - T)$;
the point $T=1$ is thermally unstable, ($\de \cool/\de T = -1$), while
$T=0$ represents the thermally stable phase. In addition to the equilibrium
points, this function also correctly mimicks the negative slope of the actual cooling 
function in the range $10^4$~K $\simlt T \simlt 10^6$~K.

Fig. 1 shows the temporal evolution of the CC front, as obtained from the 
numerical solution of eq. (2.1) with the above cooling function. 
Initially a cold cloud at $T=0$ is immersed in the hot gas at $T=1$. The
evolution of the CC front induces a thermal wave in the cold medium and a
cooling wave into the hot medium; the temperature profile is very steep due
to the strong nonlinearity of the thermal conduction (shown by the enlarged
view in Fig. 1). At about $t\simeq t_c$ 
the thermal wave propagation is inhibited by radiative losses (see FS93),    
and the cloud now acts as a finite perturbation for the unstable hot gas,
whose only possible fate is to condensate. As the CC front moves
away from the origin, a cooling wave develops ($t\simeq 2 t_c$); the 
profile of the wave becomes
stationary and its shape corresponds to a quasi-linear thermal conductivity, in
that $(\de T/ \de q)^2 \ll \vert T(\de^2 T/\de q^2)\vert$. This fact suggests to
compare the numerical results with an analytical travelling wave solution of 
eq. (2.1) obtained for linear conductivity ($\chi=const.$). 
Making the position $T(q,t)=T(\xi=q+5Ut)$, substitution into eq. (2.1) 
yields the solution

$$T(\xi)=1-{1\over (1+e^{-U\xi})^2},\eqno(2.2)$$

\noindent where $U$ is the propagation speed of the wave. 
In Fig. 1 we have fitted the curve of eq. (2.2) to the cooling wave, adjusting
the parameter $U$ to take into account that the exact solution is only
quasi-linear; the agreement is remarkable for $U=4/3=\epsilon^{-1}$. 
Of course, it is not
surprising that the value of $U$ is close to unity, since it can be seen
from dimensional analysis that it must be $U \sim \qf/t_c =1$, in our
units. The evolution depends rather weakly on the detailed shape of
$\cool$ as long as an unstable point does exist. We will use this
approximation to study the interface bewteen \lya clouds and the IGM in the
next Section. 

\bigskip
\centerline{\bf 3. Conductive/cooling model for \lya absorbers}
\medskip
To model \lya clouds as cosmological CC fronts it is necessary to 
adopt a model for the IGM. Ikeuchi \& Ostriker (1986)  (and more recently
Giroux \& Shapiro 1995) have studied in detail the thermal history of the IGM
from the era of the reheating to the present; they conclude that the most
plausible scenario requires that both photoionization and bulk mechanical
heating (\ie shocks) contribute to the heating. The IGM reionization epoch
is still uncertain, but the limits on the Gunn-Peterson effect (Webb \etal
1992) suggest that it must have occurred at $z\ge  5$. Ikeuchi \& Ostriker
show that, if the reionization occurred at $z=10$ - for example - the 
IGM entered the adiabatic expansion phase already at $z\simeq 4$.
Since we will mostly concentrate on the $z\simlt 4$ epoch, we will 
assume in what follows that the IGM is adiabatic, and therefore
temperature and pressure are $T(z)=T_0(1+z)^2$, $p(z)=p_0(1+z)^5$;
the local values are $T_0=3\times 10^4$~K and $p_0=3 \times 10^{-2}$~cm$^{-3}$~K;
we use the cosmology $\Omega=1$ and $h_{100}=0.5$; in addition $\Omega_b=0.03$. 

Since the heat
transport relies on the presence of a  hot IGM, it is likely that CC fronts 
were generated immediately after the reionization epoch. Next, they  
propagate away from the cloud with velocity decreasing rapidly 
with redshift: $U \simeq \qf/t_c \propto (1+z)^6$, assuming that IGM cooling
is dominated at early epochs by inverse Compton scattering on the microwave background.
A stationary cooling wave profile, similar to the one given in eq. (2.2),
will form after a few cooling times, as demonstrated in Fig. 1; 
this requires that the Hubble time is larger than $t_c$, a condition
satisfied for $z\simgt 4$. If reionization occurred at $z=10$, the distance
travelled by the cooling wave at $z=4$ is $d = (m_HH_0)^{-1}\int_4^{10} dz
(1+z)^{-5/2}~U n^{-1} = 60$~kpc; 
the cool material behind the wave is accreted by the parent
protogalaxy at a rate $\dot M_{in}\sim m_H n d^3 t_{H,4}^{-1} = 5\times 10^7$~
M$_\odot$~Gyr$^{-1}$, where $ t_{H,4}$ is the Hubble time at $z=4$.
As the front has reached the steady-state, its width (of 
the order of $\qf$) is regulated only by the changes of the IGM parameters 
due to Hubble expansion; since $\qf \propto (1+z)^2$, the front shrinks as
$z$ decreases (it can be shown that the interface reacts almost instantaneously
to external changes). 

The neutral hydrogen column density of the \lya cloud, $N_{HI}$, will be
in general a function both of the impact parameter, $b$, at which the interface
is intersected by the line of sight and of redshift. Assuming spherical symmetry,
with radial coordinate $r(q)$, the expression for $N_{HI}$ is

$$N_{HI}(b,z) = {2\over m_H} \int\limits_{q(b)}^{\epsilon \qf} dq~\lbrace 1 -
x[T(q,z),n(q,z)]\rbrace {r(q)\over \sqrt{r^2(q)-b^2}}.\eqno(3.1)$$
 
\noindent The hydrogen ionization fraction $x$ has been derived under the
assumption of ionization equilibrium due to photo- and collisional ionizations
and that the time variation of the diffuse UV flux is $J(z)= 10^{-21}[(1+z)/3.5]^2$
ergs~cm$^{-2}$~s$^{-1}$~Hz$^{-1}$~sr$^{-1}$. The temperature profile
follows from eq. (2.2) and $n(q)=p/kT(q)$; we have neglected geometrical 
corrections in the solution $T(q)$ due to spherical geometry. 
The lower integration limit is the Lagrangian impact parameter implicitly defined by  
$m_H^{-1}\int_0^{q(b)} {dq/n(q)}=b-d $, 
while the upper limit is the characteristic size of the
interface. The size of the central cold core is determined by the distance
$d \simeq 60$~kpc travelled by the front in the early stages of evolution; 
however, the
value of the integral depends very weakly on any reasonable choice for $d$.  
A sketch of the geometry is given in Fig. 2. Impact parameters $b \le d$ are 
not considered because of the negligible contribution of the \lya cloud to the 
central damped \lyad system.

Fig. 3 shows the curves for $N_{HI}$ from eq. (3.1) as a function of 
$q(b)$ for  different redshifts. Therefore, $q(b)=0$ corresponds in physical
units to an impact parameter $b=d$ and $q(b)=1$ corresponds to 
$b=\epsilon \lf+d$, where $\lf=\qf/m_Hn$ is the Field length in physical units;
$\lf$ depends on redshift as                           
 
$$ \lf = \left({\chi T c_p\over \cool \rho^2}\right)^{1/2}= 
333~(1+z)^{-1} {\rm ~kpc},\eqno(3.2)$$

\noindent for the adopted IGM parameters.

The hydrogen column density is a rather flat function of $q(b)$ for values
$q(b) \le 0.1$, \ie close to the central cold cloud, and decreases
roughly as $q(b)^{-4}$ for $z=4$ and less steeply for smaller values
of $z$. The values of $N_{HI}$ are in the range $10^{12} - 
3\times 10^{16}$~cm$^{-2}$ for $0 \le z\le 4$. 
An useful analytical approximation to the curves shown in Fig. 3 
between $z=0$ and $z=3$ is 

$$N_{HI}(q)=N_{HI}^0 \left[{(1+z)^5\over 1 + \left({q\over a}\right)^{1+z}}\right],
\eqno(3.3)$$

\noindent with $N_{HI}^0= 7 \times 10^{12}$~cm$^{-2}$, and $a=0.1$.

\bigskip
\centerline{\bf 4. Implications} 
\medskip 

In the present model $N_{HI}$
is a function both of redshift and of the impact parameter for the cloud: it
decreases with decreasing $z$ and with increasing $b$; the variation range
is $10^{12}{\rm cm}^{-2} \simlt N_{HI} \simlt 3 \times 10^{16}{\rm ~cm}^{-2}$
for $0 < z < 4$. Therefore, below $N_{HI}= 10^{12}$ we expect a turnoff in
the distribution of the clouds; this prediction awaits an observational test
from oncoming high sensitivity observations with the Keck telescope.

The temperature is a function of the depth inside the cloud along the
line of sight (see Fig. 2), with radial dependence as described by eq.(2.2). Its
range is bracketed by the temperature of the cold cloud (here assumed to
be $T_c=10^4$~K) and the temperature of the IGM, $T(z)$ (roughly 
$10^5-7\times 10^5$~K). For a given \lya cloud, the temperature is anticorrelated
with $N_{HI}$, since the inner regions are colder than the external ones.
However, since this cold material is far from hydrostatic
equilibrium in the gravitational field of the parent protogalaxy,
it will be accelerated towards the center acquiring bulk velocity that
could reverse the sign of the $T - N_{HI}$ correlation. This conclusion
must be substantiated by fully hydrodynamical calculations.

An interesting feature of the model is the presence of a natural scale
for the size of the clouds, the Field length $\lf$  (eq. [3.2]). Thus,
the size of \lya clouds is constrained directly by physical arguments. 
For the adopted model of the IGM, typical transverse sizes are $\sim 
50$~kpc and $\sim 150$~kpc, for $z=4$ and $z=1$, respectively. 
Hence, for a constant number of absorbers per comoving volume, the 
probability to detect a \lya cloud in a quasar pair should 
be higher at low redshifts {\it if the signal-to-noise ratio were infinite} 
(see below).

The expected redshift distribution of absorbers can be calculated
as follows. The number of absorbers between $z$ and $z+dz$ is 

$$dN = N_a c dt \int\limits_0^{b_{m}} db 2\pi b={\pi N_0 c\over H_0} (1+z)^{1/2}
b_m^2 dz,\eqno(4.1),$$ 

\noindent where $N_a=N_0(1+z)^3$ is the density of absorbers. 
The maximum impact parameter, $b_m$, is determined by the 
observational threshold for detection of \lya absorption: $b_m=b(N_{HI}^{th})$.
For example, from Fig. 3, at $z=3$ and for $N_{HI}^{th}=10^{13}$~cm$^{-2}$, which 
represents a typical threshold for current observations, the maximum
value of $q(b)$ sampled is $\sim 0.4$ (there is an almost linear relation 
between $q(b)$ and $b$). For the previous value of $N_{HI}^{th}$, 
a fit to the numerical results gives $b_m\propto \lf (1+z)^\alpha$. 
In $1 \le z \le 4$ the best fit value is $\alpha = 1.4$. 
For a higher value of $N_{HI}^{th}$, the corresponding value of 
$\alpha$ increases, \ie the distribution becomes steeper. 
Upon substitution in eq. (4.1), with $\alpha = 1.4$, we obtain
${dN/dz}\propto (1+z)^{1.3}$.
 
\noindent The index of the distribution is slightly smaller than the one found
by Bechtold (1994) ($\gamma = 1.32 \pm 0.24$) for a sample of lines complete to
equivalent width $W_{th}=0.16$~\AA. According to her curves of growth, depending
on the adopted Doppler parameter, this corresponds to 
$6\times 10^{13}{\rm cm}^{-2} \simlt N_{HI} \simlt 10^{14}{\rm cm}^{-2}$, about 
an order of magnitude larger than $N_{HI}^{th}$. Since Bechtold
has also noted a decreasing trend for $\gamma$ with decreasing $W_{th}$, 
it is reasonable to expect
that for lower $W_{th}$ the distribution flattens to a value similar to the 
one we have found. However, this physical effect can be difficult to disentangle
from spurious line blending effects (Trevese \etal 1992).

The column density distribution for the absorbers is

$${dN\over dN_{HI}}= 2 \pi N_0 \int\limits_0^{z_m} dz~b(N_{HI},z){db\over dq}(N_{HI},z) 
{dq\over dN_{HI}}(N_{HI},z) (1+z)^3 {dt\over dz}.\eqno(4.2)$$
Using the definition of the Lagrangian coordinate $q$ to eliminate $b$ and eq. (3.3)
to eliminate $q$, we get  
$dN/dN_{HI} \propto N_{HI}^{-2}  f(N_{HI}, z_m).$

Numerical integration shows that the function $f(N_{HI}, z_m)$ depends very
weakly on $N_{HI}$ and can be approximated by a power-law with index $-0.1$.
Therefore, from eq. (4.3) we find that the column density distribution 
for the \lya clouds is also a power-law with index $\beta= -2.1$. 
The value of $\beta$ is independent of $z_m$ and somewhat higher than the
one ($\beta= -1.4$) obtained by Dobrzycki \& Bechtold (1996).

Two additional consequences of our model can be outlined:
i) \lya clouds should be quite often associated
with damped \lyad clouds, from which they are originated, at least at high $z$.
At low $z$, damped \lyad systems can be destroyed by violent episodes of
star formation (Salpeter 1995) and isolated ``fossile'' \lya clouds remain; 
ii) metals abundances at the level detected by Tytler (1995) could be easily
reproduced by our model  if part of the \lya gas is coming from
a metal-rich protogalaxy, or if  the IGM is polluted by metals associated with  
the same blast waves responsible for its bulk heating.

\smallskip
\centerline{\bf Acknowledgments}                       
\smallskip
It is a pleasure to thank E. Giallongo, P. Shapiro and particularly D. Bowen
for stimulating discussions.
Yu.S. acknowledges the hospitality of the Osservatorio Astrofisico di Arcetri 
and the partial support from ESO grant B-06-013 and Russian Foundation for 
Basic Research grant 94-02-05016-a.

\vfill\eject

 
\parindent=0pc
\parskip=6pt
\centerline{\bf References}
\vskip 2pc


\ref Bechtold, J. 1994, ApJS, 91, 1 

\ref Bechtold, J., Crotts, A. P. S., Duncan, R. C, \& Fang, Y. 1994, ApJ, 437,
L83

\ref Carswell, R. F., Lanzetta, K. M., Parnell, H. C., \& Webb, J. K. 1991, ApJ,
371, 36

\ref  Dinshaw, N., Impey, C. D., Foltz, C. B., Weymann, R. J. \& Chaffee, F. H.
1994, 437, L87

\ref Dobrzycki, A. \& Bechtold, J. 1996, ApJ, 457, 102

\ref Ferrara, A. \& Shchekinov, Yu. 1993, ApJ, 417, 595 (FS93)

\ref Ferrara, A. \& Shchekinov, Yu. 1996, MNRAS, submitted

\ref Field, G. B. 1965, ApJ, 142, 531


\ref Giallongo, E., Cristiani, S., Fontana, A. \& Trevese, D. 1993, ApJ, 416,
137

\ref Giroux, M. \& Shapiro , P. R. 1995, ApJS, in press 

\ref Ikeuchi, S. \& Ostriker, J. P. 1986, ApJ, 301, 522


\ref Pettini, M., Hunstead, R. W., Smith, L. J. \& Mar, D. P. 1990, MNRAS, 246, 545

\ref Rauch, M., Carswell, R. F., Webb, J. K., \& Weymann, R. J. 1993, MNRAS,
260, 589

\ref Salpeter, E. E. 1995, in The Physics of the Interstellar and Intergalactic
Medium, PASP Series, Vol. 80,  eds. A. Ferrara, C. F. McKee, C. Heiles, P. R.
Shapiro, (PASP: San Francisco), in press



\ref Trevese, D. Giallongo, E. \& Camurani, L. 1992, ApJ, 398, 491

\ref Tytler, D. 1995, in ``Proceedings of the ESO QSO Absorption Line Workshop'', 
in press

\ref Webb, J. K., Barcons, X., Carswell, R. F. \& Parnell, H. C. 1992, MNRAS,
255, 319
\vfill\eject
 
 
\vskip 3truecm
{\bf Figure 1} 
Evolution of the structure of a conductive/cooling front; time is in units of
the cooling time, $t_c$, in the hot gas; temperature is normalized to the
temperature in the hot gas. Solid curves refer to (from right to left)
$t = 0, 0.5,1,2,3,10$; the dotted curve shows the analytical approximation for
$t=10$. The inner panel shows an enlargement of the region close to the origin.

\vskip 3truecm
{\bf Figure 2} 
Sketch of the geometry used for the conductive/cooling model.

\vskip 3truecm
{\bf Figure 3} 
Distribution of the hydrogen column density for a \lya cloud as a function
of the Lagrangian impact parameter $q(b)$ (see text) for different redshifts
$z= 4,3,2,1,0$ from the uppermost to lowest curves; $q(b)$ is in units of 
the Lagrangian Field length at the relevant redshift.
\vfill\eject
\bye
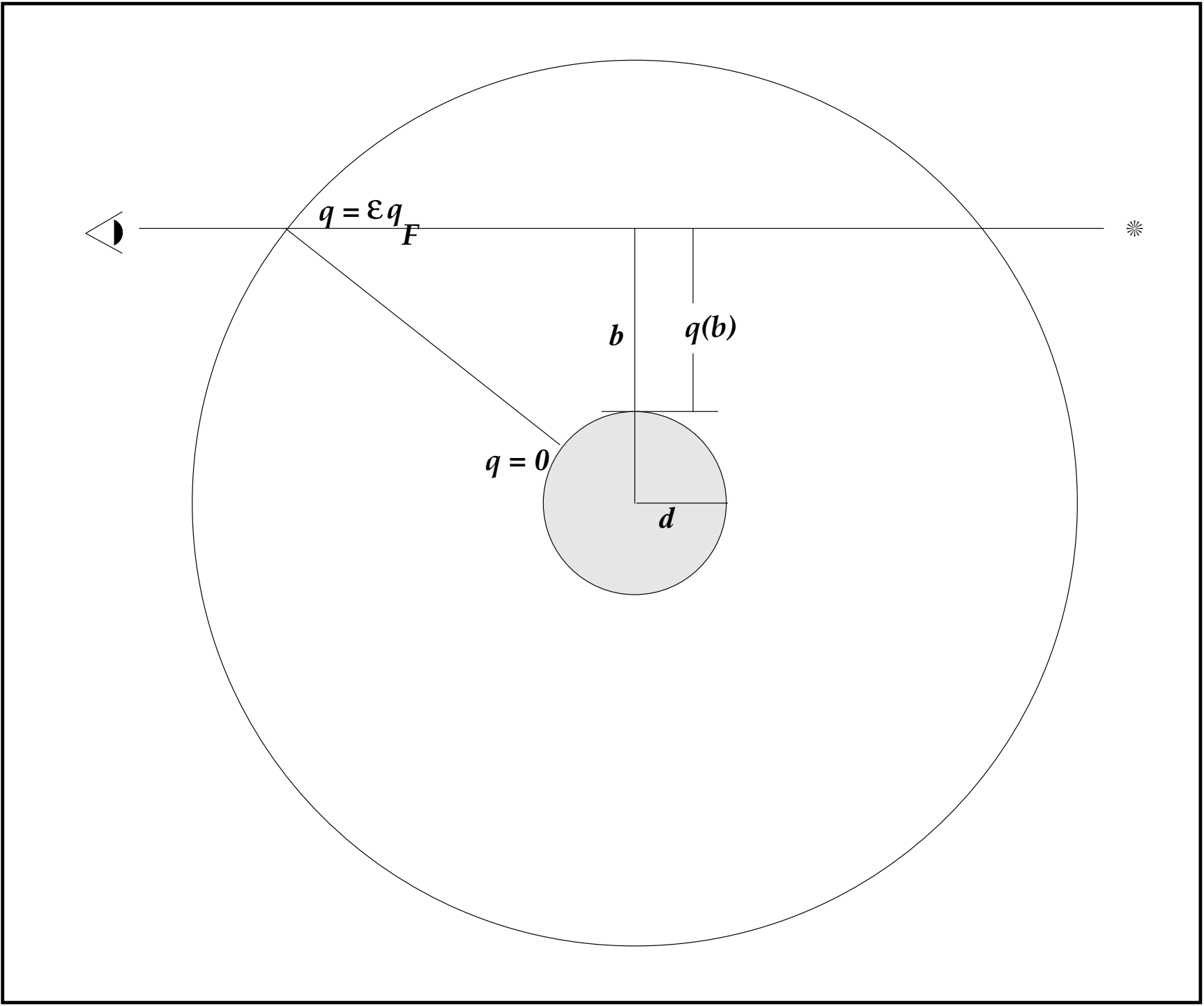
\end